\documentclass[twocolumn,9pt]{article} 

\usepackage[square,numbers,sort&compress,comma]{natbib}

\usepackage{amsmath}
\usepackage{amssymb}
\usepackage{caption}
\usepackage{graphicx}
\usepackage{latexsym}
\usepackage{times}
\usepackage[pagewise]{lineno}
\usepackage{hyperref}
\usepackage{enumitem}

\topmargin - 12pt 
\renewenvironment{abstract}%
              {
               \small
               {\bfseries \abstractname}
               \par
               \vspace{10pt}
              }

\renewcommand\abstractname{Abstract}

\newcommand{\nomenclature}
              [1]
              {
               \bgroup
               \flushleft
               \small\bf
               #1
               \par
               \egroup
              }

\renewcommand{\section}
              [1]
              {
               \bgroup
               \flushleft
               \small\bf
               \refstepcounter{section}
               \arabic{section}. #1
               \par
               \egroup
              }

\renewcommand{\subsection}
              [1]
              {
               \bgroup
               \flushleft
               \small\em
               \refstepcounter{subsection}
               \arabic{section}.
               \arabic{subsection}. #1
               \par
               \egroup
              }

\renewcommand{\subsubsection}
              [1]
              {
               \bgroup
               \flushleft
               \small\em
               \refstepcounter{subsubsection}
               \arabic{section}.
               \arabic{subsection}.
               \arabic{subsubsection}. #1
               \par
               \egroup
              }

  \newcommand{\acknowledgement}
              [1]
              {
               \bgroup
               \flushleft
               \small\bf
               #1
               \par
               \egroup
              }

  \newcommand{\sectionbib}
              [1]
              {
               \bgroup
               \flushleft
               \small\bf
               #1
               \par
               \egroup
              }

\usepackage[usenames]{color}
\definecolor{myred}{rgb}{0.8,0,0}
\newcommand{\added}[1]{#1}

\begin{document}



\small
\baselineskip 10pt

\setcounter{page}{1}
\title{\LARGE \bf Uncovering flame physics with machine learning: application to the reaction rate in hydrogen flames}

\author{{\large Antonio Attili$^{a,*}$, Ludovico Nista$^b$, Tommaso Baffetti$^b$, Geveen Arumapperuma$^a$, } \\
{Sofiane Al Kassar$^a$,  Lukas Berger$^{c,d}$, Christoph D. K. Schumann$^e$, }\\
{Temistocle Grenga$^f$, Heinz Pitsch$^b$}\\[10pt]
{\footnotesize \em$^a$School of Engineering, Institute for Multiscale Thermofluids, University of Edinburgh, Edinburgh, EH9 3FD, United Kingdom.}\\[-5pt]
{\footnotesize \em$^b$Institute for Combustion Technology, RWTH Aachen University, Aachen, 52056, Germany.}\\[-5pt]
{\footnotesize \em$^c$Thermo and Fluid Dynamics (FLOW), Faculty of Engineering, Vrije Universiteit Brussel, Brussels 1050, Belgium}\\[-5pt]
{\footnotesize \em$^d$Brussels Institute for Thermal-Fluid Systems and Clean Energy (BRITE), VUB-ULB, Brussels 1050, Belgium}\\[-5pt]
{\footnotesize \em$^e$Department of Engineering, University of Cambridge, Cambridge, CB2 1PZ, United Kingdom.}\\[-5pt]
{\footnotesize \em$^f$Faculty of Engineering and Physical Sciences, University of Southampton, Southampton, SO17 1BJ, United Kingdom.}\\[-5pt]
}

\date{}  

\twocolumn[\begin{@twocolumnfalse}
\maketitle
\rule{\textwidth}{0.5pt}
\vspace{-5pt}

\begin{abstract} 

Convolutional Neural Networks (CNNs) are used as analytical tools to investigate the relationship between the progress variable field and the local chemical source term in lean premixed hydrogen flames. Rather than employing machine learning for modelling, CNNs are leveraged to analyse the physical information contained in spatially resolved fields from direct numerical simulations. CNNs are particularly well suited for this task because they exploit spatial correlations and multi-scale structures, allowing an assessment of how spatial organisation influences the source term. 
The analysis demonstrates that the progress variable based on water, $C_{\rm H_2O}$, contains all the information required to accurately parametrise the chemical source term whereas $C_{\rm H_2}$ alone does not. 
\added{The inclusion of the mixture fraction $Z$ improves the accuracy of the latter but provides no significant additional information to the CNN when the spatial field of $C_{\rm H2O}$ is used as input.}
The same behaviour is observed in both a laminar thermodiffusively unstable flame and a turbulent slot-jet hydrogen flame, indicating that $C_{\rm H_2O}$ is a robust single variable for parametrisation. A complementary scale analysis in the laminar case shows that spatial features extending over at least two laminar flame thicknesses are required to reconstruct the source term accurately, thereby identifying the characteristic scale in the progress variable field that carries this information. These results demonstrate how machine learning can uncover physical dependencies that remain hidden to classical statistical analyses.

\end{abstract}

\vspace{10pt}

{\bf Novelty and significance statement}

\vspace{10pt}
This work introduces Convolutional Neural Networks (CNNs) as analysis tools to investigate the physical information embedded in flame fields, rather than as modelling instruments. Exploiting their ability to extract spatial correlations from high-fidelity data, CNNs are used to test whether the progress variable field alone contains sufficient information to describe the local chemical source term in lean premixed hydrogen flames. This provides a novel machine-learning framework for analysing combustion physics and source-term parametrisation. The results show that the progress variable based on water contains all the information required to accurately parametrise the source term in both laminar and turbulent flames, whereas the hydrogen-based progress variable does not. In addition, a minimum spatial extent of about two flame thicknesses is identified as the relevant scale for accurate parametrisation. These findings highlight machine learning as a powerful diagnostic tool complementary to classical statistical analyses.

\vspace{5pt}
\parbox{1.0\textwidth}{\footnotesize {\em Keywords:} hydrogen flames; convolutional neural networks; source-term parametrisation; machine learning.}
\rule{\textwidth}{0.5pt}
*Corresponding author.
\vspace{5pt}
\end{@twocolumnfalse}] 

\section{Introduction\label{sec:introduction}} \addvspace{10pt}

Lean premixed hydrogen flames are highly susceptible to thermodiffusive effects. Under laminar conditions, they exhibit complex spatial patterns and can experience several-fold increases in flame speed~\citep{berger2022intrinsic_part1,berger2022intrinsic_part2,al_kassar_efficient_2024}, while in turbulent regimes these effects interact synergistically with the flow field, producing even stronger departures from classical premixed-flame behaviour~\citep{BERGER2022112254,howarth2023thermodiffusively,berger2024effects}. A direct consequence is that the local source term and local burning rate become strongly non-uniform, with pronounced variations along the flame surface.
This has major implications for modelling, since a parametrisation of the chemical source term is often required in combustion models. 
Understanding the spatial variability of the chemical source term and identifying the variables that control it are therefore important for both physical interpretation and the development of reliable reduced-order models.
It is well known that parametrising the source term in hydrogen flames in terms of local thermochemical quantities requires at least two variables~\citep{regele2013two,berger2025combustion,bottler2024can}. This contrasts with methane flames~\citep{pitsch2006large,attili2021turbulent}, for which the source term can often be described accurately as a function of a single progress variable $C$. 
Several choices for the second control variable have been proposed, with slightly different levels of accuracy.
A widely used and effective option is the mixture fraction $Z$~\citep{regele2013two}, while a second progress variable also provides good performance~\citep{remiddi2024data}.

An alternative possibility is to replace the second variable with spatial information derived directly from the progress variable field itself~\citep{bottler2024can}. This idea is motivated by the known correlation between chemical source term and local progress-variable curvature~\citep{olguin2024closure}, since curvature contains partial information about the local spatial organisation of the flame. Although this approach improves the parametrisation, its accuracy remains limited~\citep{berger2022intrinsic_part2,bottler2024can}, suggesting that the relevant spatial features extend beyond those captured by purely local geometric quantities.
The objectives of this paper are therefore: 
\begin{enumerate}[label=\roman*)]
    \item to determine whether the progress variable field alone contains sufficient information to describe the source term
    \item to compare different progress-variable definitions, such as those based on ${\rm H_2}$ and ${\rm H_2O}$
    \item to identify the minimum spatial extent required for accurate parametrisation.
\end{enumerate}

\added{
Several data-analysis techniques can be used to extract information from reacting-flow datasets. Proper orthogonal decomposition (POD)~\cite{bizon2010pod}, for example, identifies dominant energetic modes and provides a low-dimensional representation of the flow field, while correlation-based approaches and conditional statistics can be used to quantify dependencies between selected variables. These methods have provided substantial insight into turbulent and reacting flows. However, they generally require either a linear decomposition, a predefined set of conditioning variables, or an explicitly chosen form of the correlation to be analysed. In principle, conditional statistics could also be extended to include non-local information by conditioning on the values of a scalar field at several neighbouring points. In practice, however, this would lead to a very high-dimensional conditional space, requiring prohibitively large amounts of data to obtain converged statistics. As a consequence, these approaches are less suited to determining whether a spatially distributed scalar field contains sufficient information to reconstruct a target quantity through nonlinear and non-local dependencies.
}

To address these objectives, machine learning is employed in this work. Differently from its more common use as a modelling tool, it is used here as a means of analysing the underlying flame physics.
This represents a novel use of deep learning, where neural networks are used to reveal and quantify physical relationships rather than to construct closure models. In the present case, the questions of interest cannot be readily formulated in terms of simple statistical quantities, while a direct assessment through conventional statistical analyses would require prohibitively large datasets. Machine learning offers a practical alternative, enabling the exploration of complex high-dimensional relationships and providing direct access to the information content embedded in the spatial structure of the fields.

\begin{figure}[h!]
\centering
\includegraphics[width=\columnwidth]{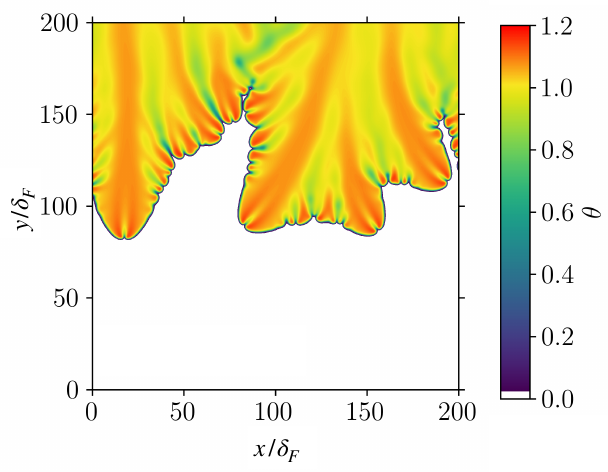}
\caption{\footnotesize Normalised temperature field in the laminar unstable hydrogen flame.}
\label{fig:2D_vis}
\end{figure}

\begin{figure}[h!]
\centering
\includegraphics[width=0.5\columnwidth]{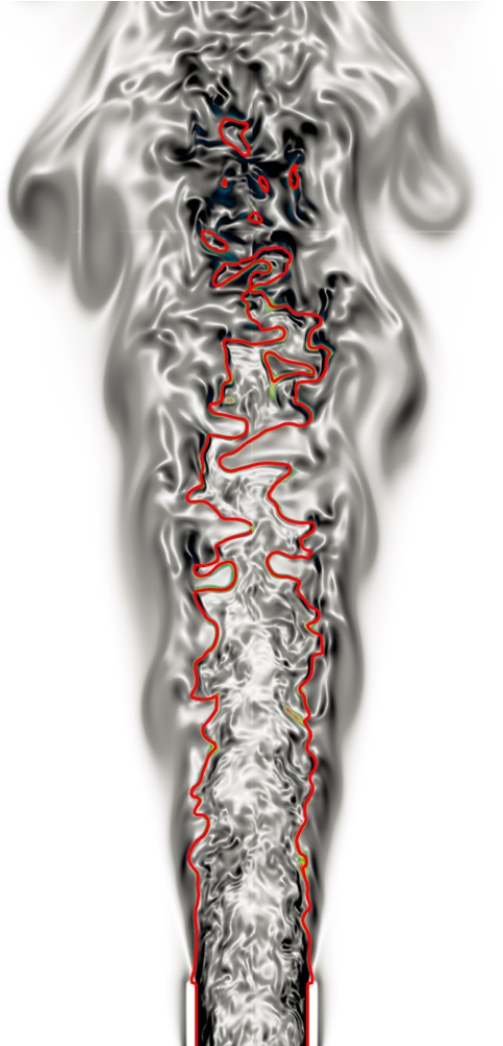}
\caption{\footnotesize Two-dimensional slice of the vorticity field in the turbulent hydrogen flame. The red line shows an isocontour of the hydrogen mass fraction, identifying the flame surface.}
\label{fig:3D_vis}
\end{figure}

\begin{figure*}[h!]
\centering
\includegraphics[width=0.8\textwidth]{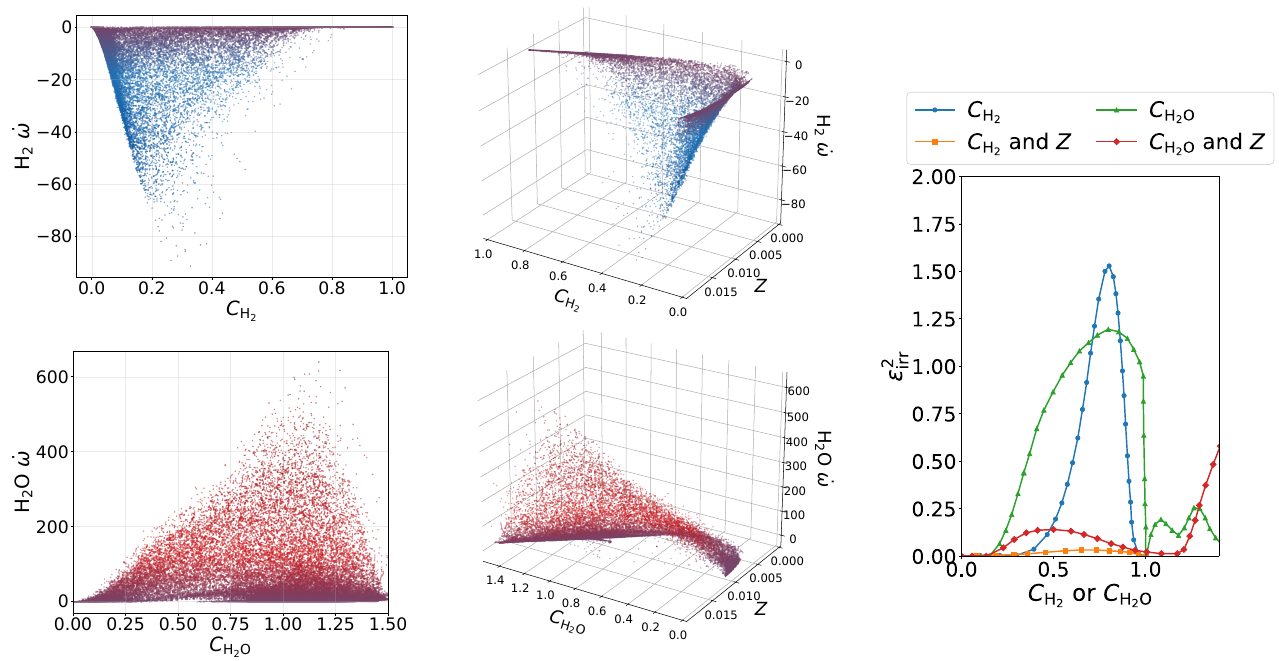}
\caption{\footnotesize Classical parametrisation of chemical source term in a lean, premixed, turbulent hydrogen flames using local variables only. Left: Manifolds in the space of progress variable (based on the hydrogen and water mass fraction) and source term. 
Centre: Manifolds in the space of progress variable, mixture fraction (based on the Bilger definition~\citep{berger2022intrinsic_part2}), and source term.
Right: Irreducible error analysis.}
\label{fig:param_H2}
\end{figure*}

\section{Datasets used in the analysis\label{sec:DNS}} \addvspace{10pt}

Data from two direct numerical simulations (DNS) of laminar and turbulent lean premixed hydrogen flames are analysed in this work. For both cases, the species, energy, and Navier-Stokes equations in the low-Mach number limit are solved using a semi-implicit finite-difference method~\citep{desjardins_high_2008}, which has been extensively applied in a variety of reacting-flow configurations~\citep{attili_effects_2016, berger_characteristic_2019, attili_turbulent_2021, BERGER2022112254}. Spatial derivatives are discretised using second-order finite differences for the momentum equation and scalar diffusive terms, while a weighted essentially non-oscillatory (WENO) scheme~\citep{liu_weighted_1994} is employed for the convective terms in the scalar equations. The chemical source terms are treated through operator splitting and integrated using the stiff ordinary differential equation solver CVODE~\citep{hindmarsh_sundials_2005}.

The first configuration is a two-dimensional laminar flame freely propagating in a square domain sufficiently large to capture the fully developed nonlinear regime of the thermodiffusive instability~\citep{berger2022intrinsic_part2}. The equivalence ratio is $\phi=0.4$, the pressure is atmospheric, and the unburned-gas temperature is $T_u=298~{\rm K}$. In this configuration, the thermodiffusive instability develops naturally from the initial perturbations, altering the flame shape and propagation speed and producing strong source-term variations along the flame surface. This configuration has been used extensively in previous studies~\citep{berger2022intrinsic_part2, schneider2025flame, creta2020propagation, howarth2022empirical} and is recognised as a convenient canonical case for investigating the fundamental physics of thermodiffusively unstable hydrogen flames and testing modelling hypotheses. The domain is discretised with a uniform mesh of resolution $\Delta$, such that $\delta_f/\Delta \approx 10$, where $\delta_f$ is the thermal flame thickness, defined as $\delta_f=(T_b-T_u)/\|\nabla T\|_{\rm max}$, with $T_b$ and $T_u$ the burned- and unburned-gas temperatures. The normalised temperature field for this flame is shown in Fig.~\ref{fig:2D_vis}.

The second configuration is a turbulent jet premixed flame surrounded by a coflow of burned gases~\citep{BERGER2022112254, kassar2026turbulent}. The jet consists of a hydrogen-air mixture with equivalence ratio $\phi=0.4$ at ambient conditions ($p=1$~atm, $T_u=298$~K), with Reynolds number $Re=UH/\nu=11\,200$. Based on the inlet velocity, jet width, and laminar flame speed, the nominal Karlovitz number is in the range $Ka=20$--$40$, where $Ka=\delta_f^2/\eta^2$ and $\eta=(\overline{\nu}^3/\tilde{\epsilon})^{1/4}$ is the Kolmogorov scale, with $\tilde{\epsilon}$ the Favre-averaged energy dissipation and $\overline{\nu}=\overline{\mu/\rho}$ the ensemble-averaged viscosity. The domain is periodic in the spanwise direction ($z$), with open outlet conditions in the streamwise direction ($x$) and slip conditions imposed at the lateral boundaries ($y$). Turbulent inlet fluctuations are generated from an auxiliary simulation of fully developed channel flow. A uniform mesh is used in all three directions, with $\Delta/\eta\approx1$ and $\delta_f/\Delta \approx 10$. The total number of grid points is approximately $1.4$ billion. 
A two-dimensional slice of the vorticity field for the slot-burner configuration is shown in Fig.~\ref{fig:3D_vis}.

\begin{figure*}[h!]
\centering
\includegraphics[width=\textwidth]{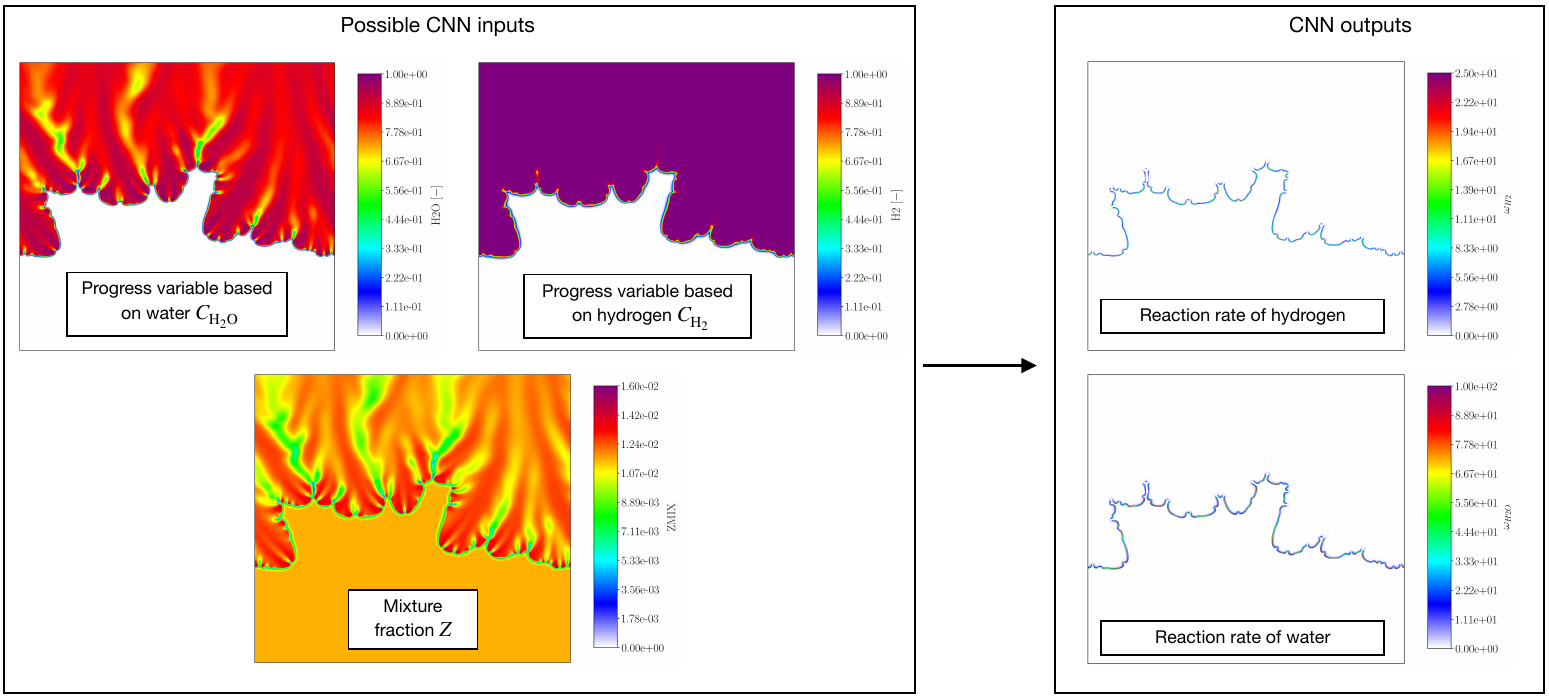}
\caption{\footnotesize Input and output fields considered for the different CNN trainings and performance evaluations in the unstable laminar hydrogen flame. The reaction rates are normalised with the peak values in the corresponding one-dimensional unstretched laminar flame.}
\label{fig:in_out}
\end{figure*}

\section{Local parametrisation of chemical source term in premixed hydrogen flames\label{sec:param}} \addvspace{10pt}

\added{

The need for two local quantities to parametrise the source term in hydrogen flames~\citep{regele2013two,berger2025combustion,bottler2024can} is illustrated in Fig.~\ref{fig:param_H2}, which reports data from the turbulent flame DNS introduced in Sec.~\ref{sec:DNS}. The figure shows local parametrisations based on both progress-variable definitions considered in this work, $C_{\rm H2}$ and $C_{\rm H2O}$, either alone or in combination with the mixture fraction $Z$. The scatter plots in the space of progress variable $C$, mixture fraction $Z$, and source term reveal well-defined two-dimensional manifolds, indicating that the parametrisation in terms of $(C,Z)$ is appropriate for both progress-variable definitions. However, the source term varies significantly at fixed $C$ as $Z$ changes, demonstrating that a parametrisation based solely on local values of the progress variable is inadequate.


A quantitative assessment is provided by the irreducible error associated with parametrisations based on $C$ alone and on $(C,Z)$~\citep{berger2025combustion}. The irreducible error measures the residual variance of the target quantity around its conditional mean for a given set of parameters and therefore represents the minimum error achievable with that parametrisation~\citep{berger2018numerically,berger2022intrinsic_part2}. As shown in Fig.~\ref{fig:param_H2}, the irreducible error remains large when only $C_{\rm H2}$ or only $C_{\rm H2O}$ is used, whereas it is strongly reduced when $Z$ is included. 
This confirms that, for both progress-variable definitions, the local value of $C$ alone is not sufficient to accurately parametrise the source term, and that mixture fraction is needed when a purely local parametrisation is considered.

One of the central questions addressed in the present work is whether this limitation of local parametrisations can be overcome by exploiting the spatial structure of the progress variable field. The underlying idea is that spatial features may contain additional information about the local flame state, allowing the source term to be reconstructed without explicitly introducing a second thermochemical variable.

}

\begin{figure*}[h!]
\centering
\includegraphics[width=0.65\textwidth]{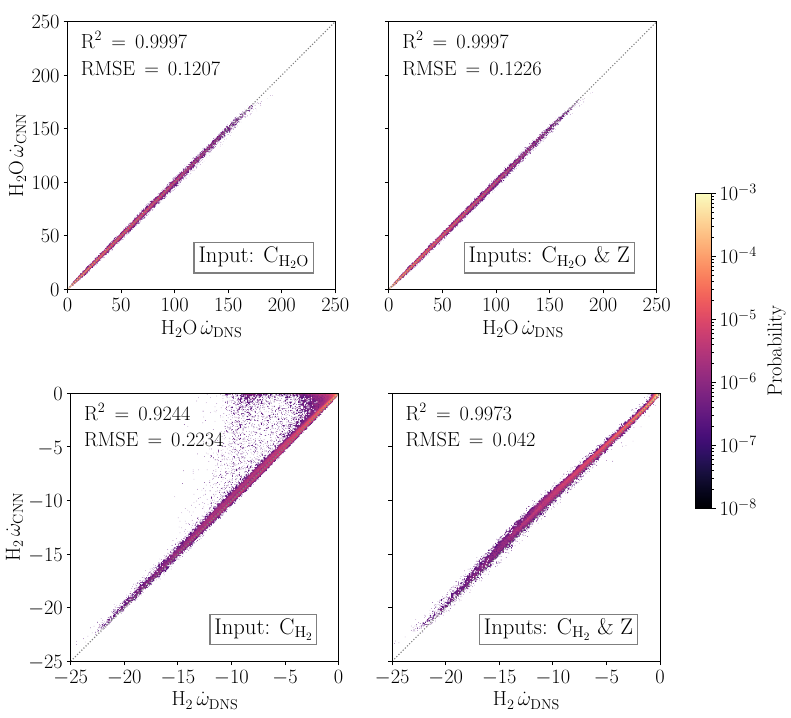}
\caption{\footnotesize Parametrisation of chemical source term in a laminar unstable flame using CNNs and different combinations of inputs. jPDFs of CNN predictions and ground truth DNS values for different combinations of inputs. Correlation coefficient and root mean square errors are also shown in each plot.}
\label{fig:fields_jpdf}
\end{figure*}

\begin{figure*}[h!]
\centering
\includegraphics[width=\textwidth]{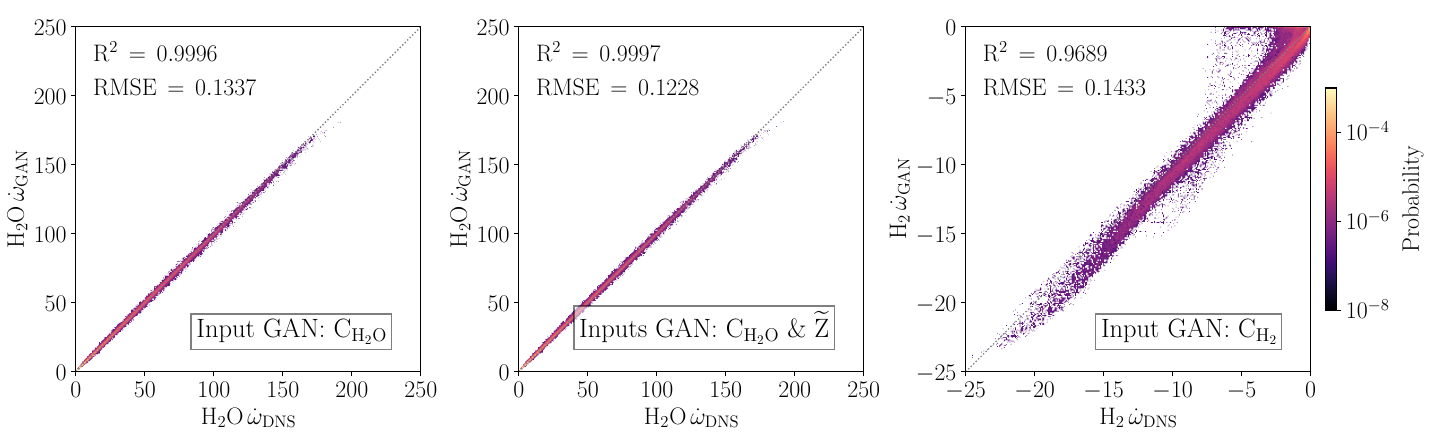}
\caption{\footnotesize Results obtained using a Generative Adversarial Network (GAN)~\citep{Nista2024_parallel,nista2024influence} for source-term reconstruction in the laminar flame, included to assess the robustness of the conclusions with respect to the machine-learning architecture employed (CNN versus GAN).}
\label{fig:GAN}
\end{figure*}

\section{CNNs as analysis tools\label{sec:cnn}} \addvspace{10pt}

\begin{figure*}[h!]
\centering
\includegraphics[width=0.65\textwidth]{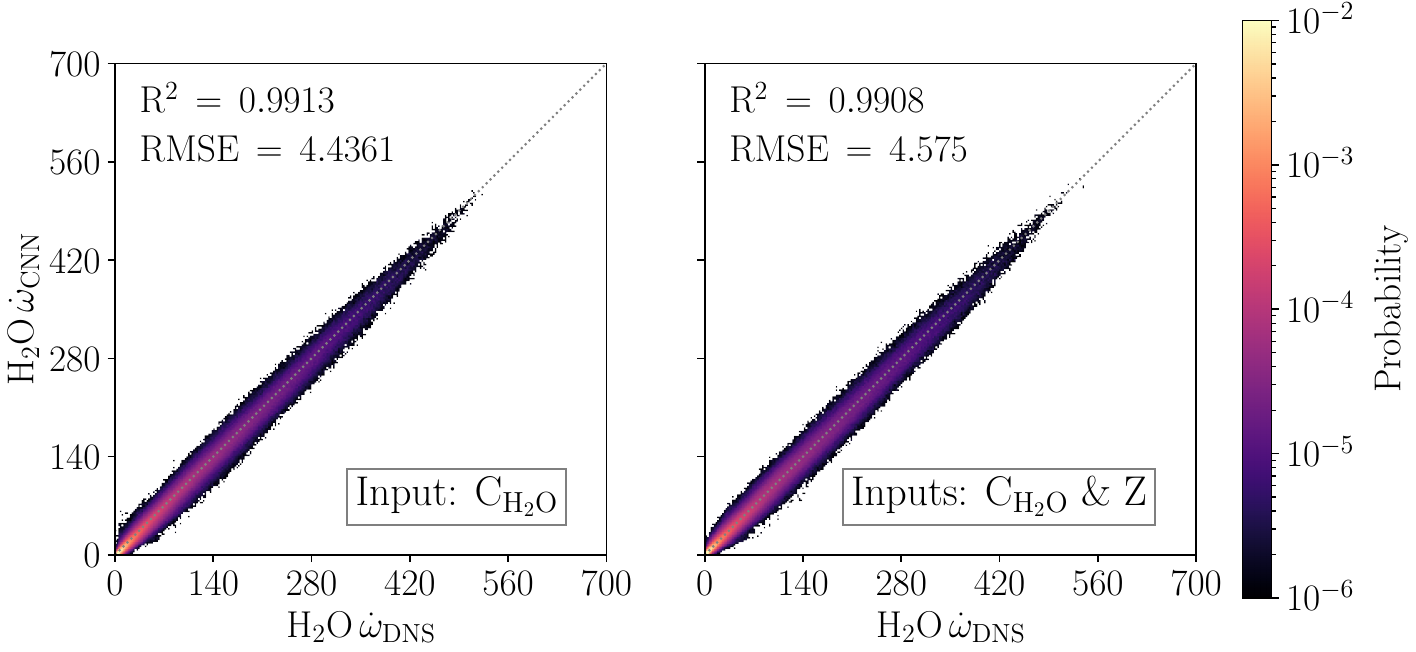}
\caption{\footnotesize Parametrisation of chemical source term in a turbulent flame using CNNs with progress variable only (left) and with the combination of progress variable and mixture fraction (right).}
\label{fig:turb}
\end{figure*}

Convolutional Neural Networks (CNNs) are employed here as a data-driven framework to examine the relationship between the spatial structure of the progress variable field and the local chemical source term in hydrogen flames. CNNs are particularly well suited for analysing spatially structured data, as they exploit local correlations through convolutional filters that progressively extract multi-scale features from one or more input fields. Among the available architectures, the U-Net structure~\cite{ronneberger2015u} is especially effective for field-to-field regression tasks and has already demonstrated excellent performance in combustion applications~\cite{lapeyre2019training,seltz2019direct, arumapperuma2025extrapolation} and it will be used in the analysis presented. Its encoder-decoder structure, combined with skip connections, allows local and global information to be processed simultaneously: as the receptive field increases through the network, progressively more complex features are learned while preserving spatial detail.
\added{
The U-Net employed here follows the same architecture and training strategy as in Ref.~\cite{arumapperuma2025extrapolation}. It is a fully convolutional encoder--decoder network composed of downsampling and upsampling blocks connected by skip connections. Each downsampling block applies convolutional operations, batch normalisation, and nonlinear activation, while the upsampling path reconstructs the output field at the original resolution. The network is trained in a supervised patch-to-patch manner: spatial blocks are extracted from the DNS fields, the selected input quantities are provided to the CNN, and the corresponding DNS source-term fields are used as targets. The trainable parameters are optimized by minimizing a mean-squared-error loss, while an independent validation dataset is used during training to monitor convergence and possible overfitting. The final accuracy is then evaluated on test fields that are not used during the optimization procedure.
}

The network’s performance, quantified through the reconstruction error between predicted and DNS source terms, is interpreted as a quantitative indicator of the physical information contained in the input fields. A low reconstruction error indicates that the selected inputs contain sufficient information to determine the source term, whereas larger errors reveal missing information in the parametrisation. To assess this systematically, the CNN is trained and evaluated using different combinations of the input fields shown in Fig.~\ref{fig:in_out}. First, only the progress variable field is used for the CNN training, in order to test whether its spatial structure alone is sufficient to reconstruct the source term. The mixture fraction $Z$ is then added as a second input, and different progress-variable definitions are considered. Finally, the network is trained using progressively larger sub-domains of the progress variable field, making it possible to quantify how spatial extent affects predictive accuracy and to identify the minimum characteristic length scale below which the parametrisation deteriorates.
\added{
Other combinations of inputs could be considered, for example two different progress variables or additional species fields. However, the main objective of the present analysis is to determine whether a single suitably chosen progress variable can provide sufficient information to infer the chemical source term. From this perspective, testing many additional combinations would not provide significant additional insight into the central question addressed in the paper.
}

For selected cases, the analysis was also repeated using a different neural-network architecture, a Generative Adversarial Network (GAN)~\citep{Nista2024_parallel,nista2024influence}, in order to verify that the observed trends were not specific to the chosen machine-learning framework.
In this approach, the generator network employs the same architecture as the CNN described above (a U-Net structure), while the discriminator is a deconvolutional network~\citep{Nista2024_parallel,nista2024influence}. 

\section{Flame analysis leveraging CNNs\label{sec:results}} \addvspace{10pt}
\subsection{Parametrisation of source term\label{sec:param_cnn}} \addvspace{10pt}

\begin{figure*}[h!]
\centering
\includegraphics[width=0.65\textwidth]{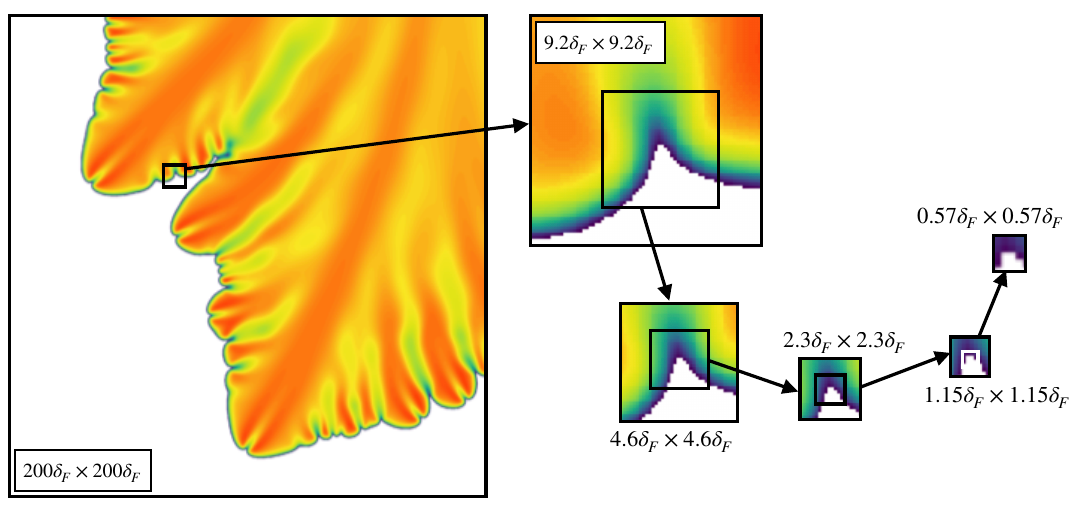}
\caption{\footnotesize Scale analysis of the chemical source term. Different box sizes used to train the different CNN models.}
\label{fig:scale_vis}
\end{figure*}

\begin{figure*}[h!]
\centering
\includegraphics[width=0.65\textwidth]{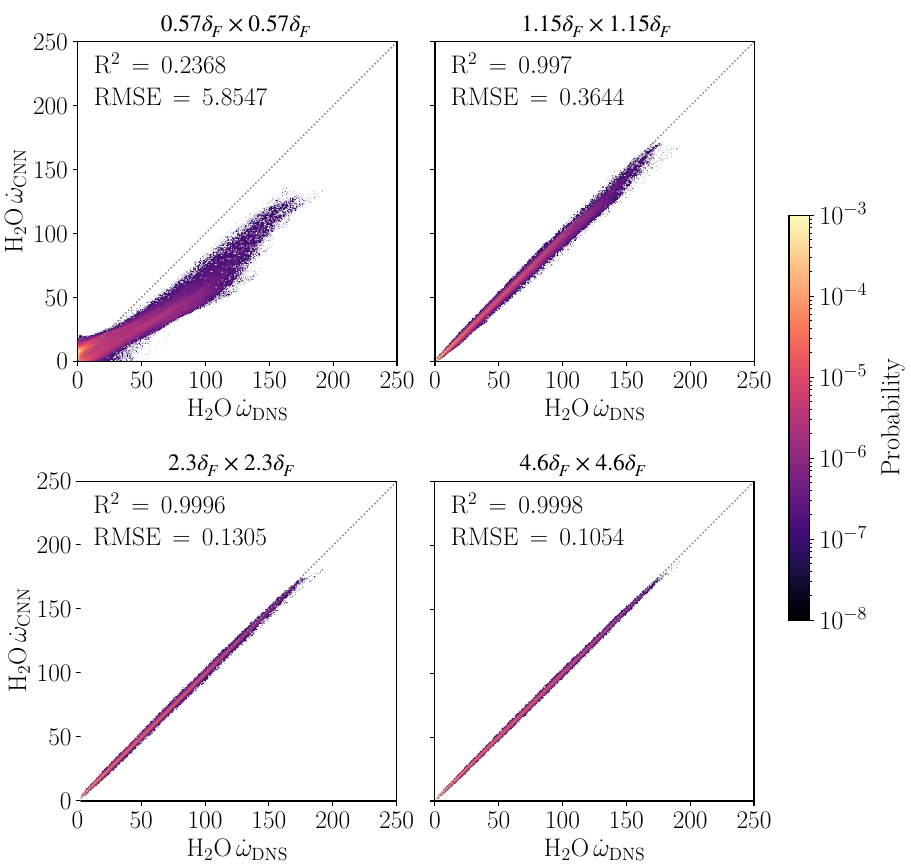}
\caption{\footnotesize Scale analysis of the source term. jPDF of CNN predictions and ground truth DNS for various box sizes.}
\label{fig:scale_jpd}
\end{figure*}

\begin{figure}[h!]
\centering
\includegraphics[width=\columnwidth]{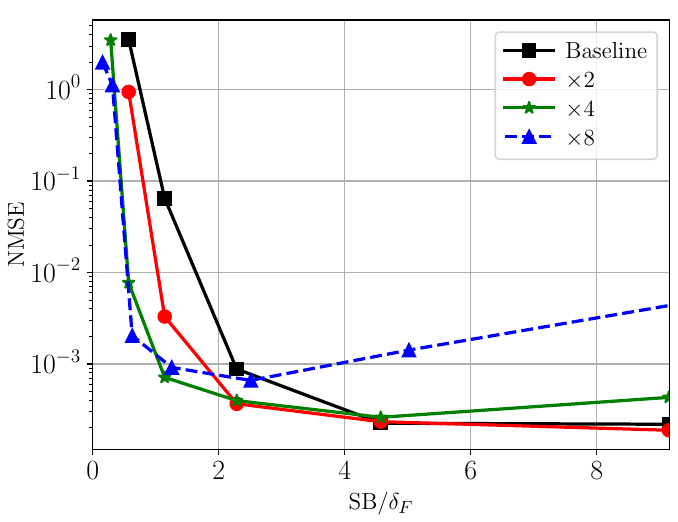}
\caption{\footnotesize Scale analysis of the source term. Error in the parametrisation for different box sizes. Different lines are for CNNs trained with data interpolated on increasingly finer meshes before training.}
\label{fig:scale_sum}
\end{figure}

The first step of the analysis consists in assessing how accurately different input fields allow the CNN to reconstruct the local source term, with the objective of identifying which quantities contain the information required for an effective parametrisation. The analysis is first carried out for the laminar flame introduced in Sec.~\ref{sec:DNS}. Four input combinations are considered (see Fig.~\ref{fig:in_out}): 
\begin{itemize}
    \item $C_{\rm H_2O}$
    \item $C_{\rm H_2O}, Z$
    \item $C_{\rm H_2}$
    \item $C_{\rm H_2}, Z$
\end{itemize}
For each case, an independent network is trained using the same architecture and optimisation procedure, and performance is quantified through the reconstruction error between the predicted and DNS source-term fields.

\added{
It is useful to note that the two progress-variable definitions have different monotonicity and boundedness properties. The progress variable based on hydrogen, $C_{\rm H2}$, is monotonic and remains bounded between 0 and 1. This property is one of the reasons why ${\rm H_2}$-based definitions are commonly used in classical flamelet models~\cite{berger2025combustion}, where the progress variable must provide a well-posed coordinate for tabulation. Conversely, the progress variable based on water, $C_{\rm H2O}$, is not strictly monotonic and may locally exceed unity. These features would be problematic in a purely local flamelet parametrisation, both numerically and because they may lead to a non-unique or ill-posed mapping between the progress variable and the thermochemical state. In the present work, however, the CNN is used primarily as an analysis tool rather than as a combustion model, and monotonicity and strict boundedness are therefore not required. The CNN does not rely on the input field as a monotonic and bounded coordinate of a tabulated manifold, but instead infers the reaction rate from the local spatial distribution of the input field. For this reason, the non-monotonicity of $C_{\rm H2O}$ and its local excursions above unity do not prevent the CNN from extracting the information needed to predict the source term.
Moreover, even if CNNs were used as combustion models, the lack of monotonicity or strict boundedness of the input field would not represent the same type of constraint as in classical flamelet models. The CNN learns a mapping from the spatial distribution of the input field to the reaction-rate output. Therefore, the network can exploit the spatial structure of the ${\rm H_2O}$ field, even if the value of $C_{\rm H2O}$ is not monotonic and is not everywhere bounded by unity.
}

Figure~\ref{fig:fields_jpdf} reports the results in the form of joint probability density functions (jPDFs) between CNN predictions and corresponding DNS values for the four input combinations. The closer the probability density is concentrated along the diagonal, the more accurate the reconstruction of the source-term field. Clear differences emerge among the different inputs. When the progress variable based on water, $C_{\rm H_2O}$, is used as the sole input, the agreement is nearly perfect, with only negligible deviations from the diagonal.
\added{
Adding the mixture fraction $Z$ does not produce any visible improvement, indicating that the spatial field of $C_{\rm H2O}$ already contains the dominant information required to reconstruct the source term. 
}

A markedly different behaviour is observed when the progress variable based on hydrogen, $C_{\rm H_2}$, is used. In this case, the jPDF exhibits a broader spread around the diagonal, showing that the source term cannot be reconstructed with the same level of accuracy. This indicates that $C_{\rm H_2}$ does not fully encode the information needed for parametrisation of its source term when considered alone. However, when the mixture fraction $Z$ is added as a second input, the reconstruction improves substantially and the agreement returns to a level comparable to that obtained with $C_{\rm H_2O}$.

\added{
The superior performance of $C_{\rm H_2O}$ can be understood by considering its spatial structure within the flame. As shown in Fig.~\ref{fig:in_out}, the $C_{\rm H_2O}$ field preserves a richer representation of flame features. In addition, $C_{\rm H_2O}$ exhibits a stronger correlation with $Z$, meaning that part of the information normally provided by $Z$ is already embedded in the spatial distribution of $C_{\rm H_2O}$. By contrast, $C_{\rm H_2}$ misses part of this information, making the explicit inclusion of $Z$ necessary to recover an accurate parametrisation.
}

\added{
It is important to emphasize that this result does not imply a simple local correlation of the form $\dot{\omega}=f(C_{\rm H2O})$. The input to the CNN is not only the pointwise value of the progress variable, but its spatial distribution over a finite region. The correlation learned by the network is therefore non-local and high-dimensional, involving the value of $C_{\rm H2O}$ at the target point as well as in the surrounding flame structure. This distinction resolves the apparent contrast with the local parametrisation discussed in Sec.~\ref{sec:param}. In a purely local description, the source term is expressed as a function of a small number of pointwise variables, and the mixture fraction is required because the local value of the progress variable alone does not uniquely determine the thermochemical state. In the CNN analysis, instead, the surrounding spatial field provides additional information on the flame structure, transport history, and mixing state. 

It is also worth noting that reproducing this analysis with conditional statistics would require conditioning on the values of $C_{\rm H2O}$ at the target point and at many neighbouring points, leading to an extremely high-dimensional conditional space that is not practically accessible.
}
\added{
A comparison with a strictly local machine-learning model, such as a multilayer perceptron or a $1\times1$ CNN, would provide another way of testing the role of spatial information. However, such models use only pointwise input values and are therefore subject to the same information limitation as the local parametrisations shown in Fig.~\ref{fig:param_H2}. Their best achievable performance is bounded by the irreducible error associated with the corresponding local conditional mean~\cite{berger2018numerically}. Since this error remains large when only the local value of $C_{\rm H2}$ or $C_{\rm H2O}$ is used, the improved performance obtained by the CNN with $C_{\rm H2O}$ must originate from the spatial information contained in the surrounding.
}

\added{
Additional tests were performed to verify that these conclusions are not restricted to the prediction of the source term associated with the species used to define the progress variable. These results are not shown here because they do not provide additional insight beyond the trends discussed above.
In particular, using $C_{\rm H2O}$ as input to predict the ${\rm H_2}$ source term leads to the same conclusions: the prediction remains accurate and the addition of mixture fraction does not significantly improve the result. Conversely, when the ${\rm H_2}$-based progress variable is used to parametrize other quantities, the prediction quality deteriorates unless mixture fraction is also provided as an input. 
}

To verify that these conclusions are not specific to the CNN architecture, selected laminar cases were also analysed using a Generative Adversarial Network (GAN)~\citep{Nista2024_parallel,nista2024influence}. 
GANs differ from CNN training as they involve the joint training of a generator and a discriminator network. The resulting adversarial optimisation introduces learning dynamics that differ from standard regression based solely on predefined loss functions, and can lead to different reconstruction characteristics.
Figure~\ref{fig:GAN} shows the corresponding jPDFs for the cases where $C_{\rm H_2O}$, $(C_{\rm H_2O},Z)$, and $C_{\rm H_2}$ are used as input. The GAN reproduces the same trends observed with the CNN: an almost perfect reconstruction is obtained when $C_{\rm H_2O}$ is used, the addition of $Z$ does not produce any noticeable improvement, and a significantly lower accuracy is observed when only $C_{\rm H_2}$ is provided. The level of discrepancy relative to the DNS data remains essentially unchanged, indicating that both the superior performance of $C_{\rm H_2O}$ and the negligible contribution of $Z$ are not influenced by the specific network or learning strategy used.
This provides additional confidence that the observed hierarchy among the different parametrisations reflects intrinsic physical relationships between the input fields and the source term, rather than artefacts introduced by the machine-learning architecture employed.

The same analysis is then extended to the turbulent slot-jet hydrogen flame introduced in Sec.~\ref{sec:DNS}~\citep{BERGER2022112254}. This extension is important because, unlike the laminar case, turbulence introduces a broad range of interacting spatial scales, additional flame wrinkling, and enhanced fluctuations of the local thermochemical state, all of which could in principle modify the information required for an accurate parametrisation of the source term.
Figure~\ref{fig:turb} compares the jPDFs between CNN predictions and DNS source-term data obtained using $C_{\rm H_2O}$ alone and using $(C_{\rm H_2O}, Z)$ as input. In both cases, the agreement remains very good, with the probability density strongly concentrated along the diagonal. The reconstruction is slightly less accurate than in the laminar flame, which likely reflects the increased complexity of the local flame dynamics under turbulent strain. However, the two jPDFs remain essentially indistinguishable, indicating that the inclusion of $Z$ does not provide any significant improvement in predictive accuracy.
This demonstrates that the information embedded in the spatial structure of $C_{\rm H_2O}$ remains sufficient to reconstruct the local source term even under fully turbulent conditions. Although turbulence introduces additional variability, the information carried by $C_{\rm H_2O}$ alone is equivalent to that contained in $(C_{\rm H_2O},Z)$, showing that the dominant physical relationship identified in the laminar flame is preserved. This confirms that $C_{\rm H_2O}$ acts as a robust single variable for source-term parametrisation across markedly different flame regimes.

These observations highlight the unique capability of machine-learning-based analysis to extract physical relationships directly from data, revealing dependencies that are difficult to access through classical statistical approaches, such as the complete information content of $C_{\rm H_2O}$ for describing the local source term.

\subsection{Scale analysis of the chemical source term in the laminar flame\label{sec:scale}} \addvspace{10pt}

/Having established, for the laminar flame, that $C_{\rm H_2O}$ contains all the information required to describe the source term, the next step is to identify the characteristic length scale of the spatial features that encode this information. 
A series of trainings was then performed for the same CNN architecture using input data in sub-domains (``boxes'') of different sizes (SB), ranging from approximately $0.2$ to $10\,\delta_F$, where $\delta_F$ is the laminar flame thermal thickness. 
Figure~\ref{fig:scale_vis} illustrates examples of these boxes superimposed on the flame field.
By analysing the performance of the network for different sizes of the input field, it is possible to identify the minimum size required for a good parametrisation.

Figure~\ref{fig:scale_jpd} reports selected jPDFs obtained for different box sizes. A clear transition is observed: for box sizes larger than approximately $2\,\delta_F$, the predictions remain highly accurate, with probability density concentrated close to the diagonal, whereas for smaller boxes the reconstruction quality deteriorates rapidly. This indicates that the source term cannot be determined from purely local information and that spatial features extending over at least two flame thicknesses are required to recover the relevant physical content embedded in the progress-variable field.

A more quantitative view is provided in Fig.~\ref{fig:scale_sum}, which shows the reconstruction error as a function of box size. The error remains low for large boxes but increases sharply below approximately $2\,\delta_F$, confirming that this scale represents a lower bound for accurate parametrisation. 
This characteristic scale is consistent with the spatial extent over which thermodiffusive structures develop along the flame 
front~\citep{berger_characteristic_2019}.

To ensure that this behaviour is not affected by numerical resolution, the analysis was repeated using data interpolated onto progressively finer meshes before training. This procedure guarantees that even the smallest boxes contain a sufficient number of grid points and avoids artefacts associated with very limited spatial sampling. The consistency of the resulting curves confirms that the observed transition is robust.
For the finest interpolated dataset, a degradation of prediction accuracy is observed for the largest box sizes. This behaviour is attributed to convergence difficulties during CNN training, a known limitation when the number of input points becomes very large~\citep{bengio2012practical}. Since this effect is absent in the other cases, it does not affect the overall interpretation of the results.

Overall, the analysis shows that the progress-variable field must be sampled over a spatial extent of at least two laminar flame thicknesses to contain sufficient information for accurate reconstruction of the local source term. Extending the same scale analysis to turbulent flames would require a substantially larger database including different Reynolds and Karlovitz numbers, in order to separate the influence of turbulence from that of thermochemical structure. This lies beyond the scope of the present work and is left for future investigation.

\section{Conclusions\label{sec:conc}} \addvspace{10pt}

This work employed a CNN-based architecture, the U-Net framework, as analytical tools to investigate the information content of progress-variable fields in lean premixed hydrogen flames and their relation to the local source term. Unlike conventional applications of machine learning aimed at prediction or reduced-order modelling, the networks were used here as diagnostic instruments to quantify whether selected input fields contain sufficient physical information to reconstruct the source-term distribution.

The analysis showed that the progress variable based on water, $C_{\rm H_2O}$, contains all the information required to accurately parametrise the source term in both laminar thermodiffusively unstable and turbulent slot-jet hydrogen flames. By contrast, the progress variable based on hydrogen, $C_{\rm H_2}$, does not provide sufficient information when used alone, and requires the addition of the mixture fraction $Z$ to recover comparable accuracy. These results indicate that the spatial structure of $C_{\rm H_2O}$ inherently embeds the information usually supplied by a second thermochemical variable.
A complementary scale analysis performed for the laminar flame demonstrated that spatial features extending over at least two laminar flame thicknesses are necessary for accurate reconstruction of the source term. This identifies a characteristic minimum length scale over which the relevant physical information is distributed, showing that the source term cannot be determined from purely local values of the progress variable even when the optimal definition is used.

Overall, the study demonstrates that machine-learning-based analysis can reveal physically meaningful relationships directly from high-fidelity combustion data, uncovering dependencies that are difficult to access through classical statistical approaches. The methodology introduced here offers a general framework that can be extended to other combustion problems where the information content of scalar fields and the identification of relevant spatial scales remain open questions.

\acknowledgement{CRediT authorship contribution statement} \addvspace{10pt}

{\bf  AA}: designed research, performed analysis, wrote paper, supervision.
{\bf  LN}: performed analysis, discussion, wrote paper, code implementation.
{\bf  TB}: performed analysis, discussion, code implementation.
{\bf  GA}: performed analysis, discussion, paper revision.
{\bf  SAK}: discussion, paper revision.
{\bf  LB}: discussion, paper revision.
{\bf  CDK}: discussion, paper revision.
{\bf  TG}: supervision, discussion, and paper revision.
{\bf  HP}: supervision, discussion, and paper revision.

\acknowledgement{Declaration of competing interest} \addvspace{10pt}


The authors declare that they have no known competing financial interests or personal relationships that could have appeared to influence the work reported in this paper.

\acknowledgement{Acknowledgments} \addvspace{10pt}


L.N. and H.P acknowledge the German Federal Ministry of Education and Research (BMBF) and the state of North Rhine-Westphalia for supporting this work as part of the NHR funding and the European Union under the European Research Council Advanced Grant HYDROGENATE, Grant Agreement No. 101054894.

\footnotesize
\baselineskip 9pt

\clearpage
\thispagestyle{empty}
\bibliographystyle{proci}
\bibliography{PCI_LaTeX}


\newpage

\small
\baselineskip 10pt


\end{document}